\documentclass[preprint,showpacs,preprintnumbers,amsmath,amssymb,superscriptaddress,floatfix]{revtex4}
\usepackage{graphicx}

\begin{document}

\title{Steady State Entanglement in Cavity QED}

\author{ P. R. Rice}
\affiliation{Department of Physics, University of Maryland,
College Park, MD 20742, USA} \affiliation{ Department of Physics,
Miami University, Oxford, OH 45056, USA}
\author{J. Gea-Banacloche}
\affiliation{Department of Physics, University of Maryland,
College Park, MD 20742, USA} \affiliation{Department of Physics,
University of Arkansas, Fayetteville, AR 72701, USA}
\author{M. L. Terraciano}
\affiliation{Department of Physics, University of Maryland,
College Park, MD 20742, USA}
\author{D. L. Freimund}
\affiliation{Department of Physics, University of Maryland,
College Park, MD 20742, USA}
\author{ L. A. Orozco}
\affiliation{Department of Physics, University of Maryland,
College Park, MD 20742, USA}

\date{\today}

\begin{abstract}
We investigate steady state entanglement in an open quantum
system, specifically a single atom in a driven optical cavity with
cavity loss and spontaneous emission. The system reaches a steady
pure state when driven very weakly. Under these conditions, there
is an optimal value for atom-field coupling to maximize
entanglement, as larger coupling favors a loss port due to the
cavity enhanced spontaneous emission. We address ways to implement
measurements of  entanglement witnesses and find that normalized
cross-correlation functions are indicators of the entanglement in
the system. The magnitude of the equal time intensity-field cross correlation
between the transmitted field of the cavity and the fluorescence
intensity is proportional to the concurrence for
weak driving fields.
\end{abstract}
\maketitle

\section{INTRODUCTION}

The study of entanglement has emerged as a
central theme of quantum physics in recent years. It is driven both by fundamental
questions and by the increasing interest in
applications that go beyond the limit of classical physics.
Entanglement as a measurable quantity is a complicated subject, in
particular when the systems have multiple components. Here we
choose to study entanglement and its possible avenues of
quantification in an open quantum system. This system, the
canonical model of cavity QED \cite{berman94}, has a single atom
coupled to the mode of an optical cavity with two reservoirs or
avenues for extracting information: spontaneous emission and losses
from the cavity.

Two particles (or systems), $A$ and $B$ are said to be in an
entangled state if the wave function of the complete system does not
factorize, that is $|AB\rangle \neq|A\rangle|B\rangle$. One
consequence of this form of the wavefunction is that a measurement
on system $A$ yields information about system $B$ without any
direct interaction with system $B$.  For systems with the same dimension, in particular, a (pure) state is said to be
maximally entangled if tracing over one of the two systems, say $A$, leaves the other one in a totally mixed state; this means that one can gain complete knowledge of system $B$ by performing measurements on $A$ only. An example that is of
relevance to this work is the maximally entangled state of an atom
and a field mode,
$|\Psi\rangle=(1/\sqrt{2})\left(|1,g\rangle+|0,e\rangle\right)$
with the first index denoting the number of photons in the field
mode and the second ($e=excited, g=ground$) denoting the state of
the atom. A measurement of the state of the atom immediately tells
us the number of photons in the field mode; or a measurement of
the photon number immediately tells us the state of the atom.

The von Neumann entropy $E=-tr_A(\rho_A log_2 \rho_A)$ of the
reduced density matrix of system $A$, $\rho_A=tr_B(\rho_{AB})$
\cite{bennet96} quantifies the amount of entanglement in a given
bipartite quantum system in a pure state. For mixed states, on the other hand, although it is easy enough to define what is meant by a totally unentangled state---namely, one in which it is possible to represent the density operator as an incoherent superposition of factorizable states---quantifying the amount of entanglement in a partially entangled state is not, in general, simple.  The natural generalization of the pure-state measure indicated above, known as the entanglement of formation,
utilizes a decomposition of the quantum state $\rho=\sum_j
P_j|\psi_j\rangle\langle\psi_j |=\sum_j P_j\rho_j$, and then
defines $E=min(\sum_j P_j E_j)$ where $E_j$ is the von Neumann entropy
for the density matrix $\rho_j=|\psi_j\rangle\langle\psi_j |$, and the minimum is taken over all the possible decompositions, which is in general a very challenging task \cite{bennet96,wooters98}. As a result of this, alternative measures have been proposed,
such as the logarithmic negativity \cite{plenio05}.  It is also possible that some particular measurement scheme may result in a most natural unraveling of the density operator, in the sense of the quantum trajectories approach \cite{nha04} (especially for systems that are continually monitored), and in that case it may be physically meaningful to focus only on the entanglement of the (conditionally pure) states obtained via that particular unraveling.

One of the main purposes of this paper is to determine how much information about the atom-field entanglement in our canonical cavity QED system can be gleaned from the kinds of measurements represented by the traditional correlation functions of quantum optics.  As we shall show below, we are actually able to avoid the difficulties for mixed-state entanglement because, in the limit we are interested in, our system is, to a good approximation, in a pure state, in spite of its being an open system interacting with two reservoirs.

\section{CAVITY QED SYSTEM}
\begin{figure}
   \begin{center}
   \begin{tabular}{c}
   \includegraphics[width=14cm]{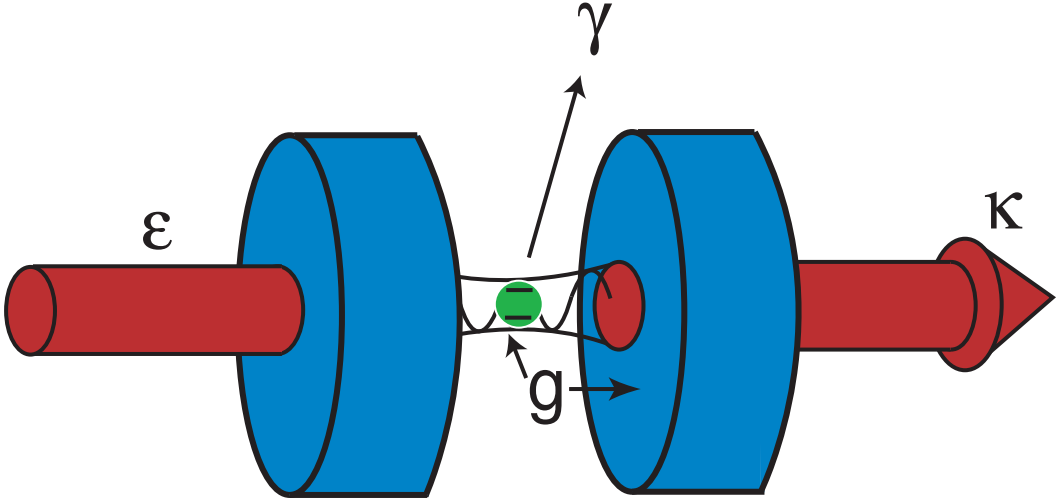}
   \end{tabular}
   \end{center}
   \caption
{Single atom in a weakly driven optical cavity. Here g is the
reversible coupling rate between the mode of the cavity and the
atom, $\kappa$ is the decay rate of the field mode of the cavity,
$\gamma$ is the spontaneous emission rate. $\epsilon$ is the
external drive (taken to be a classical field). \label{cqed}}
   \end{figure}

Fig. \ref{cqed} shows a two level atom in a driven optical
cavity. We consider a single-ended cavity, with the intracavity
field decaying via the output mirror at rate $\kappa$. The
two-level atom has a spontaneous emission rate to modes out the
sides of the cavity denoted by $\gamma$, which is generally less
than the free space Einstein $A$ coefficient. The resonant
coupling between the atom and the field mode is given by
$g=\mu_{eg}\sqrt{\omega/2\hbar\epsilon_0V}$ with $\mu_{eg}$ the
electric dipole matrix element, $\omega$ the transition frequency
and $V$ the volume of the cavity mode. The driving field is taken
to be a large classical field $\epsilon$ incident on the input
mirror, with small transmission $T_{in}$, so that the incident
flux (in photon units) inside the cavity is proportional to
$T_{in}\epsilon^2$

The quantum trajectory wave function that characterizes the system
under a non-Hermitian Hamiltonian is:
\begin{eqnarray}
|\psi _c(t)\rangle &=&\sum\limits_{n}^\infty \left(
C_{g,n}(t)e^{-iE_{g,n}t}|g,n\rangle \right.
\left. +C_{e,n}(t)e^{-iE_{e,n}t}|e,n\rangle \right) \label{psi}\\
H&=& \hbar g\;(a^\dagger \sigma _-+a\sigma _+)-i\kappa a^\dagger a
-i{{\gamma}\over2} \sigma_+\sigma_-  + i \hbar
\epsilon(a^{\dagger }-a)
\end{eqnarray}
with collapse operators
\begin{eqnarray}
{\cal A}&=&\sqrt{\kappa} a\\ {\cal
S}&=&\sqrt{{\gamma}\over2}\sigma_-.
\end{eqnarray}
associated with photons exiting the output mirror and spontaneous
emission out the side of the cavity. The indices $e(g)$ indicate
the atom in the excited (ground) state, while $n$ is the number of
photons in the mode. The energies are
$E_{e,n}=E_{g,n+1}=\hbar\omega (n+1/2)$. We have the usual
creation ($a^\dagger$) and annihilation ($a$) operators for the
field, and Pauli raising and lowering operators $\sigma_{\pm}$ for
the atom.

In the weak driving limit, the system reaches a steady-state wave
function:
\begin{equation}
|\Psi\rangle=|0g\rangle+A_{1,g}|1g\rangle+A_{0,e}|0e\rangle+A_{2,g}|2g\rangle+A_{1,e}|1e\rangle
\label{wavefunction}\end{equation} where the $A_{ij}$ are known
\cite{carmichael91,brecha99}. They are
\begin{eqnarray}
A_{1,g}&=&\alpha \\
A_{0,e}&=&\beta\\
A_{1,e}&=&\alpha\beta   q \label{twoexitation}\\
A_{2,g}&=&{\alpha}^2pq/\sqrt{2}.
\end{eqnarray}
The quantities $p$ and $q$ would be $1$ for coupled harmonic
oscillators. In cavity QED they differ from unity due to the
non-harmonic, or saturable, nature of the atom. The squares of
coefficients of single excitation $A_{1,g},~A_{0,e}$ give the
rates of detection of single photons through the output mirror or
in fluorescence (steady state), while the squares of the double
excitation coefficients $A_{1,e},~A_{2,g}$ give the rates of
detection of two photons either in coincidence (one through the
mirror, and one in fluorescence) or both out of the mirror. The
variables are
\begin{eqnarray}
\alpha&=&{\epsilon\over{\kappa(1+2C_1)}}\\
\beta&=&{{-2g}\over{\gamma}}\alpha\\
p&=&1-2C_1' \label{p}\\
q&=&{{(1+2C_1)}\over{(1+2C_1-2C^{'}_1)}}\label{q}\\
C_1&=&{{g^2}\over{\kappa \gamma}}\\
C_1^{'}&=&C_1 {{2\kappa}\over{(2\kappa+\gamma)}}
\end{eqnarray}
The one-excitation amplitudes $A_{1,g}$ and $A_{0,e}$ are
proportional to the driving field $\epsilon$; the two-excitation
amplitudes $A_{2,g}$, and $A_{1,e}$ are proportional to the square
of the driving field, $\epsilon^2$. \cite{carmichael91}. The norm
of this wave function is $||\Psi\rangle |=\sqrt{1+O(\epsilon^2)}$;
hence to lowest order in $\epsilon$, the coefficient of the vacuum
should be $(1-(1/2)O(\epsilon^2))$. The term $O(\epsilon^2)$ makes
no contribution to lowest nonzero order in $\epsilon$ for the
correlation functions or entanglement measures considered here.

The entanglement of formation for this system is calculated from
the density matrix after tracing over the field variables:
\begin{eqnarray}
\rho_{atom} &=& Tr_{field} |\Psi\rangle\langle\Psi |\\
&=& \left(
\begin{array}{cc}
  1 + A_{1,g}^2 + A_{2,g}^2 & A_{1,e}A_{1,g}+A_{0,e} \\
  A_{1,e}A_{1,g}+A_{0,e} & A_{1,e}^2 + A_{0,e}^2 \\
\end{array}
\right)
\end{eqnarray}

The eigenvalues of this matrix are, to lowest nonvanishing order,
\begin{eqnarray}
\lambda_1 &=& \left(A_{1,g} A_{0,e} -  A_{1,e} \right)^2\nonumber  \\
&=&  |A_{1,g}|^2|A_{0,e}|^2(q-1)^2\nonumber\\
&=&\left({{\epsilon}\over{\kappa}}\right)^4\xi^2\\
\lambda_2 &=& 1-\left(A_{1,g} A_{0,e} -  A_{1,e} \right)^2\nonumber\\
&=&1-\left({{\epsilon}\over{\kappa}}\right)^4\xi^2\\
\end{eqnarray}

where $q$ is defined in Eq. (\ref{q}), and we have defined
\begin{equation}
\xi={{2g}\over{\gamma (1+2C_1)^2}}(q-1)
\end{equation}

The entropy $ E = -\lambda_1 \log_2 \lambda_1 - \lambda_2
\log_2 \lambda_2 $ is then (again to lowest leading order)
\begin{eqnarray}
E&=&-\left({{\epsilon}\over{\kappa}}\right)^4\xi^2 \log_2\left[\left({{\epsilon}\over{\kappa}}\right)^4\xi^2\right] -\left(1-\left({{\epsilon}\over{\kappa}}\right)^4\xi^2\right) \log_2\left[1-\left({{\epsilon}\over{\kappa}}\right)^4\xi^2\right]\nonumber \\
&\approx&-\left({{\epsilon}\over{\kappa}}\right)^4\xi^2 \left(\log_2\left[\left({{\epsilon}\over{\kappa}}\right)^4\right]+\log_2\left[\xi^2\right]-1\right)\nonumber \\
&\approx&- \left({{\epsilon}\over{\kappa}}\right)^4
\log_2\left[\left({{\epsilon}\over{\kappa}}\right)^4\right]\xi^2.
\label{entanglement-1}
\end{eqnarray}
where we have taken the weak field limit, $\epsilon$ being the smallest
rate in the problem, so $\epsilon/\kappa \ll 1$.  The approximation (\ref{entanglement-1})
will hold provided $(\epsilon/\kappa)^2 \ll |\xi|$.

This entropy is the same as that obtained by using the density matrix for the field alone,
traced over the atomic degrees of freedon.

The concurrence, first introduced by Wooters for two qubits\cite{wooters98}, can also be
used to characterize entanglement between two quantum systems of arbitrary
dimension \cite{Chen, Uhlmann, Rungta,Albeverio}. The concurrence for our system is

\begin{eqnarray}
 {\cal C}&=&\sqrt{2(1-Tr\rho_{atom}^2)}\nonumber \\
 &=&\sqrt{4 \left(A_{1,g} A_{0,e} -  A_{1,e} \right)^2}\nonumber\\
 &=&2\left({{\epsilon}\over{\kappa}}\right)^2|\xi|
 \end{eqnarray}
To see why $|\xi|\propto |A_{1,e}-A_{0,e}A_{1,g}|$ may be a good indication
of entanglement, consider what happens if the wavefunction is a
product state. We could write
\begin{eqnarray}
|\Psi\rangle_P&=&|\psi_F\rangle\otimes|\phi_A\rangle\nonumber \\
&=&\left(D_0|0\rangle+D_1|1\rangle+D_2|2\rangle\right)\otimes\left(C_g|g\rangle+C_e|e\rangle\right)\nonumber
\\
&=&D_0C_g|0g\rangle+D_1C_g|1g\rangle+D_0C_e|0e\rangle +D_2C_g|2g\rangle+D_1C_e|1e\rangle
\end{eqnarray}
For weak excitations, the coefficient of the ground state of the
system is $D_0C_g=1$, or $C_g=D_0=1$. Then the product state is
\begin{equation}
|\Psi\rangle_P=|0g\rangle+D_1|1g\rangle+C_e|0e\rangle+D_2|2g\rangle+D_1C_e|1e\rangle
\end{equation}
Just knowing the one excitation amplitudes does not yield any information
about entanglement, as it is possible to have $A_{1,g}=D_1$ and
$A_{0,e}=C_e$. $A_{2,g}$ gives no information about entanglement,
just nonclassical effects in the field, as it only involves field
excitation. For weak fields $D_2$ is exactly $A_{2,g}$. The
entanglement shows up in the value of $A_{1,e}$; if this value
does not satisfy $A_{1,e}=D_1C_e=A_{0,e}A_{1,g}$, then it is not
possible to write the state as a product state.

In the presence of a non-zero vacuum contribution (as any real
quantum state will have), one can learn nothing about entanglement
simply by measurement of one-excitation amplitudes or
probabilities. For example, the state
$|0,g\rangle+\alpha(|1,g\rangle+|0,e\rangle) $ is entangled, but
only if one is certain that the probability amplitudes for higher
excitation are truly zero. A state of the form
$|0,g\rangle+\alpha(|1,g\rangle+|0,e\rangle) +O(\epsilon^2)$
cannot be said to be entangled without information on the relative
size of the probability amplitude $A_{1,e}$. Measurement of
one-excitation amplitudes conditioned by a previous measurement
{\it can} yield information about entanglement. This can be
accomplished by utilizing cross-correlation functions. A first
important conclusion out of this study is that a measure of the
zero time cross correlation between the atom and the field, as
well as the mean transmitted and fluorescent intensities yields a
measure of entanglement in the weak field limit.

\section{ENTANGLEMENT FOR WEAK EXCITATION \label{I}}
Equation~(\ref{entanglement-1}) of the previous section gives the
amount of entanglement in the system as a function of the one and
two excitation amplitudes. In terms of specific system parameters
the concurrence is:
\begin{equation}
{\cal C}=|2\alpha \beta (q-1)|=
\frac{16\,g^3\,{\epsilon }^2\,\kappa }
  {{\left( 2\,g^2 + \gamma \,\kappa  \right) }^2\,
    \left( 2\,g^2 + \kappa \,
       \left( \gamma  + 2\,\kappa  \right)  \right) }.
\label{entanglement}
\end{equation}

This section analyzes the sensitivity of the concurrence to the
different parameters that appear in Eq.~(\ref{entanglement}), while
trying to give physical reasons for their influence on the
entanglement. Despite the fact that the rates of decay could be
the same through the two reservoirs, spontaneous emission
($\gamma$) reduces entanglement more than cavity loss ($\kappa$).
This is due to the fact that a $\gamma$ event (spontaneous
emission) {\it must} come from the atom, while a $\kappa$ event
(cavity transmission) could come from either the drive or a photon
emitted by the atoms into the cavity mode. A spontaneous emission
event unambiguously leaves the atom in the ground state, and the
system wavefunction factorizes.

\begin{figure}
   \begin{center}
   \begin{tabular}{c}
\includegraphics[width=14cm]{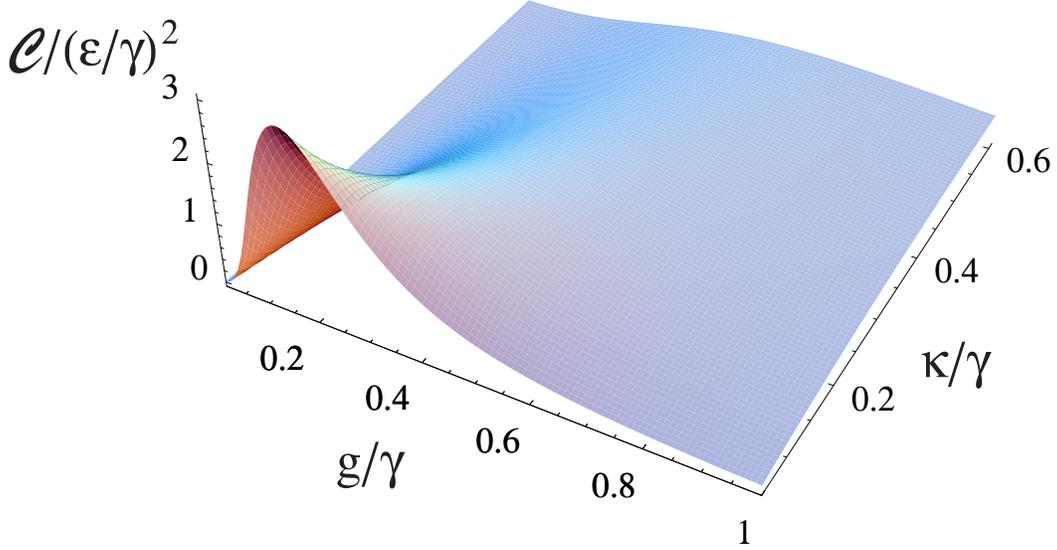}
   \end{tabular}
   \end{center}
   \caption[example]
{A plot of ${\cal C}$ scaled by $(\epsilon/\gamma)^2$ as a
function of $\kappa/\gamma$ and $g/\gamma$ for weak excitation.
\label{entan-g-gam-ka}}
   \end{figure}

Fig. \ref{entan-g-gam-ka} shows a remarkable result in the
entanglement of the system as a function of the three rates in the
problem. There is an optimal value for the coupling constant $g$
given a set of dissipation rates $\kappa,\gamma$. For many
interesting cavity QED effects, stronger coupling is generally
better, such as the enhancement of the spontaneous emission by a
factor of $1+2C_1=1+2g^2/\kappa\gamma$ (this formula strictly
holds only in the bad cavity limit $\kappa>>g,~\gamma$). However,
here increasing the coupling of the atom and field mode eventually
decreases the amount of entanglement. To explain this it is
instructive to recall that the concurrence ${\cal C}=|2 \alpha \beta
(q-1)|$, where $\alpha$ is the mean cavity field, and
$\beta=-g\alpha/\gamma$ is the mean atomic dipole. As the
coupling $g$ increases, for a fixed weak driving field $\epsilon$,
the intracavity field $\alpha=\epsilon/(\kappa+2g^2/\gamma)$
decreases. The intracavity field is the sum of the driving field
in the cavity $\epsilon/\kappa$, and the field radiated by the
atom, $(-2C_1/(1+2C_1))\epsilon/\kappa$, the minus sign resulting
from the fact that the radiated field is $\pi$ out of phase with
the driving field on resonance. We see that as $g$ and $C_1$
increase, the intracavity field decreases. This means that the
steady-state wavefunction has a larger vacuum component, and
consequently less entanglement. Another way to view this is that
the cavity enhancement of the spontaneous emission rate means a
larger loss rate for the system as the coupling increases, which
is bad for entanglement.

More formally, consider what happens if the two-excitation
amplitudes in Eq.~(\ref{wavefunction}) are arbitrarily set to zero,
which amounts to setting $q=0$ in Eq.~(\ref{entanglement}), in which
case the entanglement is only determined by the prefactor $|\alpha\beta|$.
The steady-state wave function becomes
\begin{equation}
|\psi\rangle_{ss}=|0g\rangle+\alpha(|1g\rangle-{{g}\over{\gamma}}|0e\rangle).
\label{on-wavefc}
\end{equation}
There are two interesting limits on this Eq.~(\ref{on-wavefc}) for the
parameter $f=g/\gamma$. If $f \gg 1$, the steady state
wavefunction is approximately $|\psi\rangle_{ss}=|0\rangle
(|g\rangle-f\alpha|e\rangle)$ which is a product state. Also, if
$f \ll 1$, the steady state wavefunction is approximately
$|\psi\rangle_{ss}=|g\rangle (|0\rangle+\alpha|1\rangle)$ which
again is a product state. To have entanglement between the atom
and cavity mode, we must have the parameter $f \simeq 1$, so as to
prepare a steady state wavefunction of the form
$|\psi\rangle_{ss}=|0g\rangle+\alpha(|1g\rangle-|0e\rangle)=|0g\rangle+\alpha|-\rangle$,
a mixture of the vacuum with a small entangled state component.

The decrease of the prefactor $|\alpha \beta|$ is the
dominant reason why the concurrence decreases with increasing
$g$ for large coupling. Close inspection of Fig. \ref{entan-g-gam-ka} also shows that there is an optimal
cavity loss rate $\kappa$ for entanglement for a fixed $g$ and $\gamma$.
This is a result of reaching a maximum in the population of the
states different from the vacuum (Eq.~(\ref{wavefunction})). Our
results here are consistent with the numerical results of Nha and
Carmichael \cite{nha04}.

When the system is driven off resonance, its response is typically
characterized by transmission and fluorescent spectra
\cite{gripp96a,terraciano05}. Although these are important probes of
the system, they do not, in this limit, carry information about the
entanglement, since they are derived from only the one-excitation amplitudes.

\begin{figure}
   \begin{center}
   \begin{tabular}{c}
\includegraphics[width=14cm]{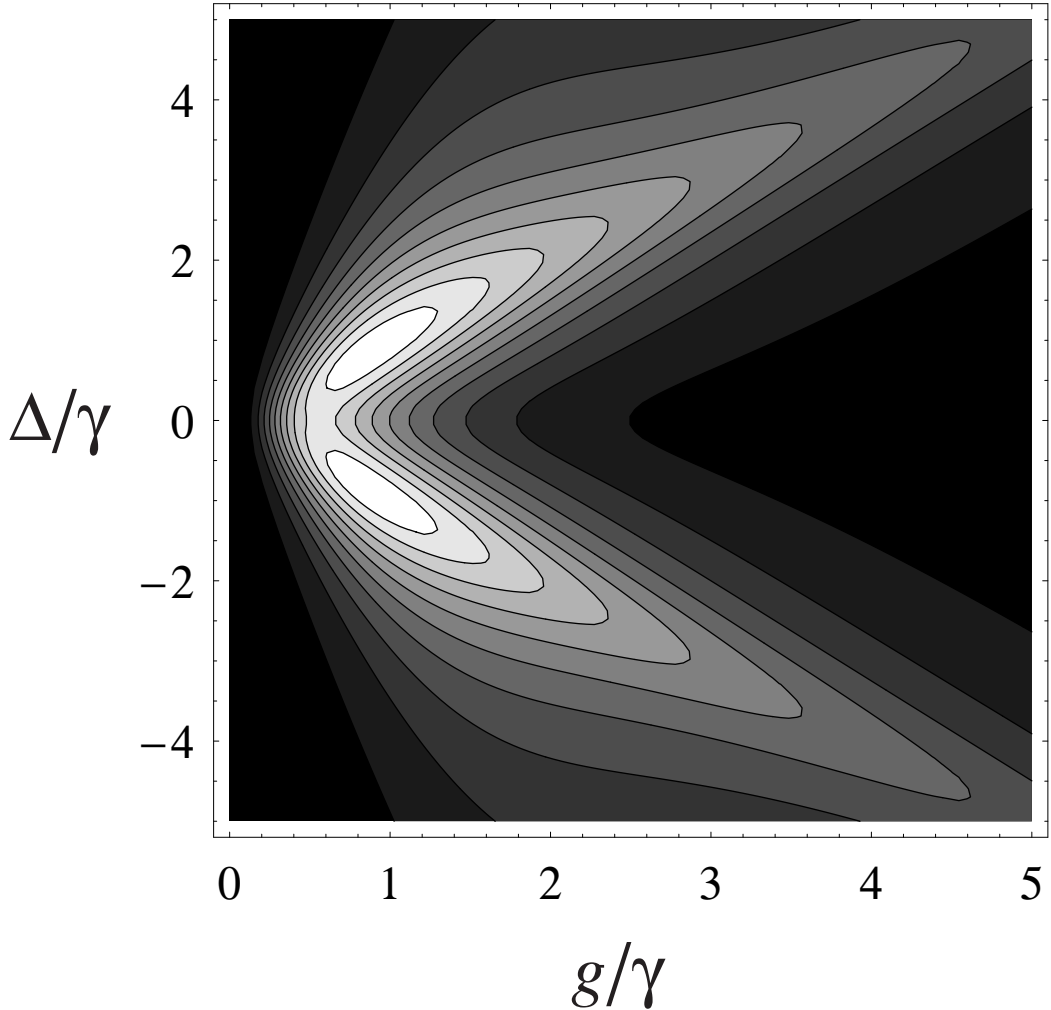}
   \end{tabular}
   \end{center}
   \caption[example]
{Contour plot of $\cal C$ as a function of
$g/\gamma$ and $\Delta/\gamma$ for $\kappa/\gamma=0.5$ \label{contour-vr}}
   \end{figure}

The concurrence as a function of the detuning of the driving laser
shows that the steady state entanglement decreases typically by a
factor of $1/\Delta^3$ for large detuning, where
$\Delta=(\omega-\omega_l)$ with $\omega$ the resonant frequency of
the atom and cavity, and $\omega_l$ the frequency of the driving
probe laser. But in the case where $g$ is larger than $\kappa$ and
$\gamma$, the response is maximized at the vacuum-Rabi peaks
\cite{carmichael89b}. Figure \ref{contour-vr} shows a contour plot
of $\cal C$ for parameters in the regime of cavity QED where the
two decay rates are similar: $2\kappa/\gamma=1.0$. The concurrence
increases with increasing $g$ on resonance up to a saddle point,
and then decreases. However the entanglement persists for
detunings on the order of $g$, the approximate location of the
vacuum-Rabi peaks in the spectra of the system.

Detuning to a vacuum-Rabi peak ($\Delta=\pm g$),
generates a steady state wave function of the form
\begin{equation}
|\psi\rangle_{ss}=|0,g\rangle+\alpha\Gamma_1(g/\gamma)|1,\pm\rangle+\alpha^2
\Gamma_2(g/\gamma)|2,\pm\rangle,
\end{equation}
where $|n,\pm\rangle=(1/\sqrt{2})(|n,g\rangle\pm|n-1,e\rangle)$ is the
$n$ photon dressed atom-field state one is tuned near and $\Gamma_1(g/\gamma)$ and $\Gamma_2(g/\gamma)$ are
functions that are maximal when $g \simeq \gamma$.
This is a
state of mainly vacuum, plus a part that has entanglement between
the atom and the cavity. It would seem that by continuing to tune
to a vacuum-Rabi peak as $g$ increases, it would be possible to
maintain the entanglement, but Fig. \ref{contour-vr} shows that this
is not the case.
Rather, as argued (for the on-resonance case) above, the crucial parameter for maximizing entanglement is $f=g/\gamma
\propto 1/\sqrt{n_{sat}}$, where $n_{sat}=\gamma^2/8g^2$ is the
saturation photon number. This is the dependence on the
nonlinearity of the atomic system. Recall that, if these were two
driven coupled harmonic oscillators, $q=1$ and there would be no
entanglement. A nonlinear interaction between the two harmonic
oscillators would be needed to entangle them, as in the signal and
idler modes in optical parametric oscillation. This nonlinear
interaction would generate two-mode squeezing, which could be
measured by homodyne detection of mode A(B) conditioned on
detection of a photon in mode B(A), just as squeezing in one mode
can be detected via conditioned homodyne detection of a mode based
on a photodetection from that mode\cite{carmichael00,foster00}.
The nonlinearity of the two-level atom is needed to generate
two-mode squeezing and entanglement between the atom and the
cavity field. Even though the driving field is weak and the atom
never nears saturation, there can only be entanglement with a
linear atom-field coupling if the atom has a nonlinear response,
as two-level atoms do.

\begin{figure}
   \begin{center}
   \begin{tabular}{c}
   \includegraphics[width=14cm]{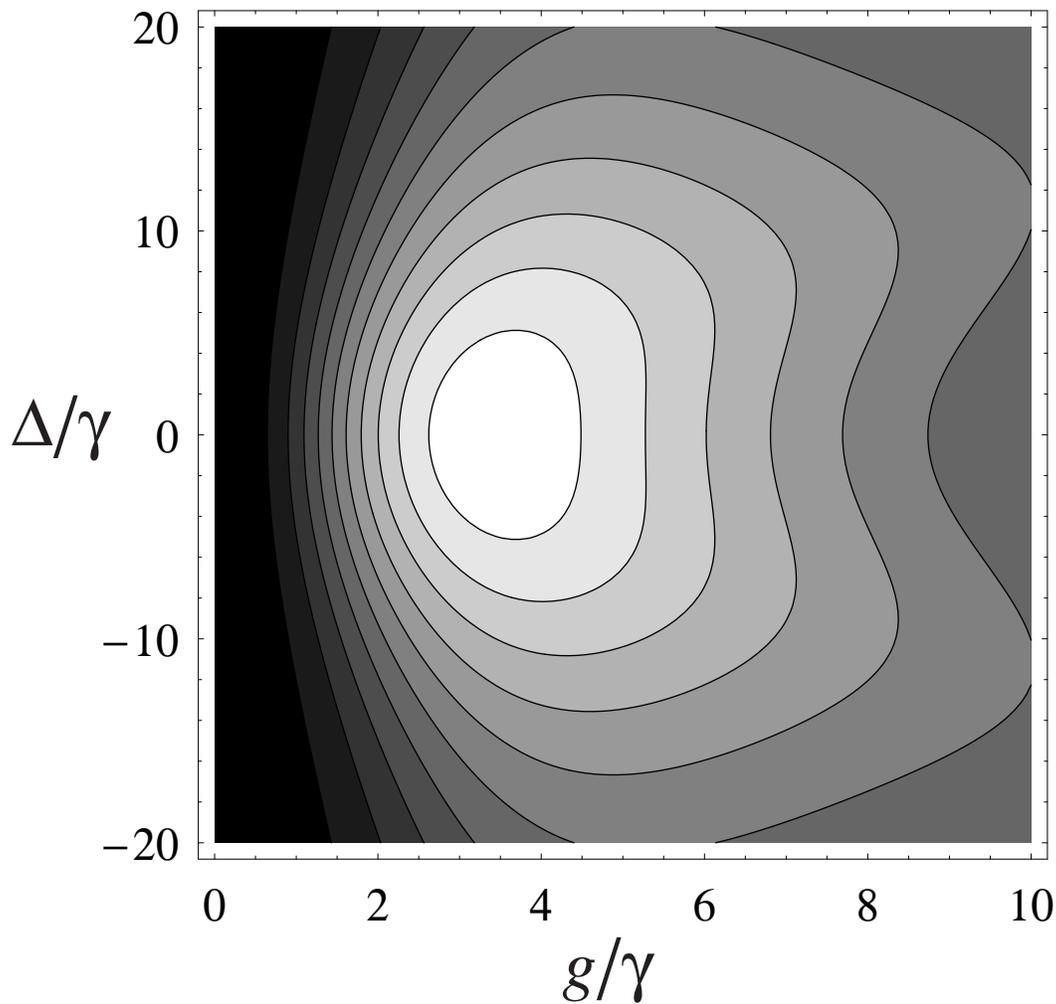}
   \end{tabular}
   \end{center}
   \caption[example]
{Contour plot of $\cal C$ as a function of $g/\gamma$ and
$\Delta/\gamma$ for $\kappa/\gamma=10$ \label{contour-nvr}}
   \end{figure}

The concurrence shows its sensitivity to
different parameters. Fig. \ref{contour-nvr} shows a contour plot
of $\cal C$ versus $g/\gamma$ and $\Delta/\gamma$ for a case
where the cavity decay rate is larger than the spontaneous
emission rate ($\kappa/\gamma=10.0$). The entanglement is largest
near $g/\gamma=4.0$, before the vacuum-Rabi splitting of the
spectrum, which does not occur in this case until $g/\gamma \sim 10.0$, at which point  the entanglement is already diminishing. The size of the
maximum concurrence decreases by increasing $\kappa/\gamma$ from
$0.5$ to $10.0$ by a factor of about 30.

\section{MEASUREMENTS OF ENTANGLEMENT WITH CORRELATION FUNCTIONS}

The calculation of entanglement leads now to the question of how
to implement measurements that give the full information in the
case of this cavity QED system under weak excitation. The previous
section shows that the concurrence is related to the rate of single
photon counts out of the cavity or in fluorescence and to the rate
of coincident counts from the cavity and fluorescence. These are
the quantities associated in quantum optics with correlation
functions, first introduced by Glauber
\cite{glauber63a,glauber63b,glauber63c,mandel95}. Generally these
correlation functions involve comparing a field (intensity) of one
mode with the field (intensity) of the same mode at a later time
(or different spatial location), with some exceptions
\cite{grangier86,kuzmich03,regelman01,berglund02,moore99,leach04}.
However, entanglement in cavity QED has two components: atom and
cavity mode. It is natural to look at cross correlations between
the cavity mode and the fluorescent light that falls in the mode
of the detector.

Consider a general cross-correlation function for two-modes of the
electromagnetic field:
\begin{equation}
G=\langle f_1(b^{\dagger},b)f_2(a^{\dagger},a)\rangle/\langle
f_1(b^{\dagger},b)\rangle\langle f_2(a^{\dagger},a)\rangle.
\end{equation}
with $f_1$ and $f_2$ well behaved functions, in the sense of a
convergent Taylor series on the Hilbert space of interest. If
$|\psi \rangle$ is a product state, the correlation function
$G(a,b)$ factorizes and then is unity. If it is {\it not} a
product state, then this will manifest itself in a non-unit value
for the normalized cross-correlation functions.

The simplest cross correlation function to consider is
$g_{TF}^{(1)}(0)$. This could be obtained by measuring the
visibility of the fringe pattern formed by interfering the
transmitted and fluorescent light. For the weakly driven
cavity-QED system, this is
\begin{eqnarray}
g^{(1)}_{TF}(0)&=&{{\langle \sigma_+a\rangle}\over{ \langle \sigma_+ \rangle }\langle a\rangle}\nonumber \\
&=&{{\alpha \beta}\over{\alpha \beta}}\nonumber \\
&=&1
\end{eqnarray}
so to lowest order, there is no information in this correlation
function about entanglement.

To obtain information about entanglement the correlation function
has to probe the two-excitation part of the state. A possibility
to do this is the intensity cross correlation:
\begin{eqnarray}
g_{TF}^{(2)}(0)&=&{{\langle \sigma_+
a^{\dagger}a\sigma_-\rangle}\over{\langle a^{\dagger}a
\rangle\langle \sigma_+ \sigma_-\rangle}}\nonumber\\
&=&{{|A_{1e}|^2}\over{|A_{1g}A_{0e}|^2}}\nonumber\\&=&q^2
\end{eqnarray}

 This normalized
correlation function is directly related to the coefficient of
double excitations (See
Eqs.~(\ref{wavefunction}),~(\ref{twoexitation}),~(\ref{q})). If
$q=1$ then $g_{TF}^{(2)}(0)=1$ and there is no entanglement; so a
non-unit value of $q$ indicates entanglement. Using second-order
intensity correlations has been proposed in the context of
entangled coherent states by Stobi{\'n}ska and W{\'o}dkiewicz
\cite{stobinska05}.

The cross correlation function $g_{TF}^{(2)}(0)$ contains
information about the average photon number {\it in coincidence
with} a measurement of the fluorescence relative to the average
photon number in the absence of any interrogation of the
fluorescence. $g_{TF}^{(2)}(0)-1=q^2-1$ is an indicator of
entanglement.

A way to measure $q$ directly utilizes a field-intensity
correlation function $h_{\theta}(\tau)$ \cite{carmichael04}, that
can be implemented as a homodyne measurement conditioned on the
detection of a fluorescent photon,
\begin{eqnarray}
h_{\theta=0}^{TF}(0)&=&{{\langle I_F E_T\rangle}\over{\langle
I_F\rangle\langle E_T\rangle}}\nonumber \\
&=& {{\langle (a^{\dagger}+a)\sigma_{+}\sigma_{-}
\rangle}\over{\langle
a^{\dagger}+a\rangle\langle \sigma_+\sigma_-\rangle}}\nonumber \\
&=&{{A_{1,e}}\over{A_{0,e}A_{1,g}}}\nonumber\\
&=&q
\end{eqnarray}

So $h_{\theta=0}^{TF}(0)-1=q-1$ is also an indicator of entanglement in this system.
What makes
this measurement possible experimentally is the conditioning that
selects only times when there is a fluctuation and the rest of the
time (when the vacuum is present) no data is collected
\cite{foster00}. For one mode, the homodyned transmitted field
conditioned by detection of a photon from that mode, is a measure
of squeezing in that mode \cite{carmichael04}. A homodyne
measurement of the transmitted field conditioned by detection of a
fluorescent photon is a measure of the two-mode squeezing, with
the cavity field and atomic dipole as the two components.
Generally, two-mode squeezing is an indicator of entanglement
between the two modes. Gea~Banacloche {\it et al.} explored this
correlation function in a different regime of cavity QED and found
it to be a witness of the dynamics of entanglement \cite{gea05}.

Non-classicality and entanglement are not necessarily
simultaneously present. For example for two oscillators one could
have $|\psi\rangle=(1/\sqrt{2})(|A,B\rangle+|B,A\rangle)$, where
$A$ and $B$ are coherent state amplitudes. In this state, there is
entanglement, but each individual mode shows no non-classical
behavior. Conversely, one can have non-classical behavior with no
entanglement, say for example the atom in the ground state and the
field in a squeezed coherent state.

There is a particular form of the Schwartz inequality that must be
satisfied for a classical field for the specific case of the
system we are considering here:
\begin{equation}
(g^{(2)}_{TF} (0) - 1)^2 \leq|(g^{(2)}_{TT} (0) - 1)(g^{(2)}_{FF}
(0) - 1)|,
\end{equation}
Here $TT$ and $FF$ denote zero delay intensity correlations for
the transmitted and fluorescent fields respectively. In the
one-atom limit, $g^{(2)}_{FF} (0) = 0$, and $g^{(2)}_{TT} (0) =
q^2p^2$, so this inequality becomes $|q^2 - 1|^2 \leq |q^2p^2 -
1|$ which depends on $q$, but also on the parameter $p$ (Eq.
(\ref{p})), which can be varied independently. There is no
one-to-one relationship between Schwarz inequality violations and
entanglement (by this measure) in this particular system.

\section{CONCLUSION}

We find that entanglement in weakly driven cavity QED  is
characterized by comparison of two-excitation probability
amplitudes to single excitation amplitudes, in particular the
amplitude involving one excitation in each subsystem. It is
necessary to have a small saturation photon number to enhance the
nonlinear response which generates a larger entanglement. But this
is true only to a point. We find the maximal entanglement for
small $\kappa$ and when $g/\gamma$ is on the order of unity. This
stems from the dual role of the coupling $g$. It couples energy
into the atom, but due to cavity enhanced spontaneous emission, it
can also channel energy out.

Increasing $\gamma$ decreases the entanglement, and this can be
explained in terms of the effect of the two decay processes on the
system. If we detect a fluorescent photon we know it has come from
the atom, and the atom is in the ground state. If we obtain a
transmitted photon, it could have been emitted from the atom into
the cavity mode, or just be a driving field photon that has passed
through the cavity without interaction with the atom. It is the
interference of these two indistinguishable processes that leads
to nonclassical effects in the transmitted field.

We have found a variety of cross-correlation functions that are
indicators, or witnesses, of entanglement in this system. One can
learn nothing about the entanglement by examining only one- or
two- excitation amplitudes separately. In particular we find that
a measurement of two-mode squeezing, or a homodyne measurement of
the transmitted field conditioned on the detection of a
fluorescence photon is directly proportional to the entanglement
calculated via the reduced von Neumann entropy. Further work
remains to generalize this approach to situations with higher
drives, but the general approach of looking at entanglement
together with the specific correlation function to measure gives
physical insight into this problem.

We would like to thank J. P. Clemens for fruitful discussions
related to the topic of this paper. This work has been supported
by NSF and NIST.



\begin{thebibliography}{10}
\newcommand{\enquote}[1]{``#1''}
\expandafter\ifx\csname url\endcsname\relax
  \def\url#1{\texttt{#1}}\fi
\expandafter\ifx\csname
urlprefix\endcsname\relax\def\urlprefix{URL }\fi
\providecommand{\eprint}[2][]{\url{#2}}

\bibitem{berman94}
P.~R. Berman, ed., \emph{Cavity Quantum Electrodynamics}, Advances
in Atomic,
  Molecular, and Optical Physics (Academic Press, Boston, 1994). Supplement 2.

\bibitem{bennet96}
C.~H. Bennett, D.~P. DiVincenzo, J.~A. Smolin, and W.~K. Wootters,
  \enquote{Mixed-state entanglement and quantum error correction,} Phys. Rev. A
  \textbf{54}, 3824-3851 (1996).

\bibitem{wooters98}
W.~K. Wootters, \enquote{Entanglement of Formation of an Arbitrary
State of Two
  Qubits,} Phys. Rev. Lett. \textbf{80}, 2245-2248 (1998).

\bibitem{plenio05}
M.~B. Plenio, \enquote{Logarithmic Negativity: A Full Entanglement
Monotone
  That is not Convex,} Phys. Rev. Lett. \textbf{95}, 090503 (2005).

\bibitem{nha04}
H.~Nha and H.~J. Carmichael, \enquote{Entanglement within the
Quantum
  Trajectory Description of Open Quantum Systems,} Phys. Rev. Lett.
  \textbf{93}, 120408 (2004).

\bibitem{carmichael91}
H.~J. Carmichael, R.~J. Brecha, and P.~R. Rice, \enquote{Quantum
Interference
  and Collapse of the Wavefunction in Cavity {QED},} Opt. Commun. \textbf{82},
  73-79 (1991).

\bibitem{brecha99}
R.~J. Brecha, P.~R. Rice, and M.~Xiao, \enquote{N Two-Level Atoms
in a Driven
  Optical Cavity: Quantum Dynamics of Forward Photon Scattering for Weak
  Incident Fields,} Phys. Rev. A \textbf{59}, 2392-2417 (1999).

\bibitem{Chen}K. Chen, S. Albeverio, and S. Fei, \enquote{Concurrence of Arbitrary Dimesional Bipartite
Quantum States,} Phys. Rev. Lett. \textbf{59}, 040504 (2005).

\bibitem{Uhlmann} A. Uhlmann, \enquote{Fidelity and Concurrence of Conjugated States,}
 Phys. Rev. A \textbf{62}, 032307 (2000).

\bibitem{Rungta}P. Rungta, V. Buzek, C. Caves, M. Hillery, and G. J. Milburn,
\enquote{Universal State Inversion and Concurrrence in Arbitrary
Dimensions,} Phys. Rev. A \textbf{64},042315 (2001).

\bibitem{Albeverio}S. Albeverio and S. Fei, \enquote{A Note On Invariants And Entanglements,}
J. Opt. B \textbf{3}, 233-327 (2001).

\bibitem{gripp96a}
J.~Gripp and L.~A. Orozco, \enquote{Evolution of the Vacuum {R}abi
Peaks in a
  Many-Atom System,} Quantum Semiclass. Opt. \textbf{8}, 823-836 (1996).

\bibitem{terraciano05}
M.~L. Terraciano, R.~Olson, D.~L. Freimund, L.~A. Orozco, and
P.~R. Rice,
  \enquote{Fluorescence spectrum into the mode of a cavity QED system,}
  arXiv.org, \textbf{quant-ph/0601064}.

\bibitem{carmichael89b}
H.~J. Carmichael, R.~J. Brecha, M.~G. Raizen, H.~J. Kimble, and
P.~R. Rice,
  \enquote{Subnatural Linewidth Averaging for Coupled Atomic and Cavity-Mode
  Oscillators,} Phys. Rev. A\textbf{40},  5516-19 (1989).

\bibitem{carmichael00}
H.~J. Carmichael, H.~M. Castro-Beltran, G.~T. Foster, and L.~A.
Orozco,
  \enquote{Giant Violations of Classical Inequalities through Conditional
  Homodyne Detection of the Quadrature Amplitudes of Light,} Phys. Rev. Lett.
  \textbf{85}, 1855-1858 (2000).

\bibitem{foster00}
G.~T. Foster, L.~A. Orozco, H.~M. Castro-Beltran, and H.~J.
Carmichael,
  \enquote{Quantum State Reduction and Conditional Time Evolution of
  Wave-Particle Correlations in Cavity {QED},} Phys. Rev. Lett. \textbf{85},
  3149-3152 (2000).

\bibitem{glauber63a}
R.~J. Glauber, \enquote{Photon Correlations,} Phys. Rev. Lett.
\textbf{10}, 84-86
  (1963).

\bibitem{glauber63b}
R.~J. Glauber, \enquote{The Quantum Theory of Optical Coherence,}
Phys. Rev.
  \textbf{130}, 2529-2539 (1963).

\bibitem{glauber63c}
R.~J. Glauber, \enquote{Coherent and Incoherent States of the
Radiation Field,}
  Phys. Rev. \textbf{131}, 2766-2788 (1963).

\bibitem{mandel95}
L.~Mandel and E.~Wolf, \emph{Optical Coherence and Quantum Optics}
(Cambridge
  University Press, New York, 1995).

\bibitem{grangier86}
P.~Grangier, G.~Roger, A.~Aspect, A.~Heidmann, and S.~Reynaud,
  \enquote{Observation of Photon Antibunching in Phase-Matched Multiatom
  Resonance Fluorescence,} Phys. Rev. Lett. \textbf{57}, 687-690 (1986).

\bibitem{kuzmich03}
A.~Kuzmich, W.~P. Bowen, A.~D. Booze, A.~Boca, C.~W. Chou, and
L.-M.~Duan and
  H.~J.~Kimble, \enquote{Generation of nonclassical photon pairs for scalable
  quantum communication with atomic ensembles,} Nature \textbf{423}, 731-734
  (2003).

\bibitem{regelman01}
D.~V. Regelman, U.~Mizrahi, D.~Gershoni, E.~Ehrenfund, W.~V.
Schoenfeld, and
  P.~M. Petroff, \enquote{Semiconductor Quantum Dot: A Quantum Light Source of
  Multicolor Photons with Tunable Statistics,} Phys. Rev. Lett. \textbf{87},
  257401 (2001).

\bibitem{berglund02}
A.~J. Berglund, A.~C. Doherty, and H.~Mabuchi, \enquote{Photon
Statistics and
  Dynamics of Fluorescence Resonance Energy Transfer,} Phys. Rev. Lett.
  \textbf{89}, 068101 (2002).

\bibitem{moore99}
M.~G. Moore and P.~Meystre, \enquote{Optical control and
entanglement of atomic
  Schr{\"o}dinger fields,} Phys. Rev. A \textbf{59}, R1754-R1757 (1999).

\bibitem{leach04}
J.~Leach, C.~E. Strimbu, and P.~Rice, \enquote{Nonclassical
cross-correlations
  of transmitted and fluorescent fields in cavity QED systems,} J. Opt. B:
  Quantum Semiclass. Opt \textbf{6}, S722-S729 (2004).

\bibitem{stobinska05}
M.~Stobi{\'n}ska and K.~W{\'o}kdiewicz, \enquote{Witnessing
entanglement with
  second-order interference,} Phys. Rev. A \textbf{71}, 032304 (2003).

\bibitem{carmichael04}
H.~J. Carmichael, G.~T. Foster, J.~E. Reiner, L.~A. Orozco, and
P.~R. Rice,
  \enquote{Intensity-Field Correlations of Non-Classical Light,} in
  \emph{Progress in Optics}, E.~Wolf, ed., vol.~46, p. 355-4-4 (Elsevier,
  Amsterdam, 2004).

\bibitem{gea05}
J.~Gea-Banacloche, P.~R. Rice, and L.~A. Orozco,
\enquote{Entangled and
  Disentangled Evolution for a Single Atom in a Driven Cavity,} Phys. Rev.
  Lett. \textbf{94}, 053603 (2005).

\end{thebibliography}
\end{document}